\documentclass[10pt,conference]{IEEEtran}
\IEEEoverridecommandlockouts
\usepackage{stmaryrd}
\usepackage{cite}
\usepackage{float}
\usepackage{amsmath,amssymb,amsfonts}
\usepackage{algorithmic}
\usepackage{graphicx} 
\usepackage[table]{xcolor}
\usepackage{textcomp}
\usepackage{lipsum}
\usepackage{xcolor}
\usepackage{booktabs}
\def\BibTeX{{\rm B\kern-.05em{\sc i\kern-.025em b}\kern-.08em
    T\kern-.1667em\lower.7ex\hbox{E}\kern-.125emX}}
\usepackage{subfigure}
\usepackage{mathtools}
\usepackage{multirow}
\usepackage{lettrine}
\usepackage[ruled,vlined,linesnumbered]{algorithm2e}
\usepackage{babel,blindtext}
\usepackage[spaces,hyphens]{url}

\usepackage{tabularx}
\usepackage[justification=centering]{caption}
\usepackage{amsthm}

\usepackage{nccmath}
\usepackage{booktabs,tikz}

\usepackage{lipsum, babel}
\usepackage{soul}

\makeatletter
\newcommand*\titleheader[1]{\gdef\@titleheader{#1}}
\AtBeginDocument{%
  \let\st@red@title\@title
  \def\@title{%
    \bgroup\normalfont\large\centering\@titleheader\par\egroup
    \vskip1.5em\st@red@title}
}
\makeatother

\title{\huge{Optimizing Information Freshness Leveraging Multi-RISs in NOMA-based IoT Networks }}
\titleheader{\small{Accepted by 2022 IEEE Globecom conference, ©2022 IEEE}}

\begin{document}

\author{Ali~Muhammad, Mohamed Elhattab, Mohamed Amine Arfaoui, and Chadi Assi \\ Concordia University, Montreal, Canada,\\ emails: $\{$muha\_a@encs, m\_elhatt@encs, m\_arfaou@encs,  assi@ciise$\}$.concordia.ca} 
\maketitle
\IEEEpeerreviewmaketitle
\begin{abstract}
This paper investigates the benefits of integrating multiple reconfigurable intelligent surfaces (RISs) in enhancing the timeliness performance of uplink Internet-of-Things (IoT) network, where IoT devices (IoTDs) upload their time-stamped status update information to a base station (BS) using non-orthogonal multiple access (NOMA). Accounting to the  potential unreliable wireless channels due to the impurities of the propagation environments, such as deep fading, blockages, etc., multiple RISs are deployed in the considered IoT network to mitigate the propagation-induced impairments, to enhance the quality of the wireless links, and to ensure that the required freshness of information is achieved. In this setup, an optimization problem has been formulated to minimize the average sum Age of Information (AoI) by optimizing the transmit power of the IoTDs, the IoTDs clustering policy, and the RISs configurations. The formulated problem ends up to be a mixed-integer non-convex problem. In order to tackle this challenge, the RISs configurations are first obtained  by adopting a semi-definite relaxation (SDR) approach. Then, the joint power allocation and user-clustering problem is solved using the concept of bi-level optimization, where the original problem is decomposed into an outer IoTDs clustering problem and an inner power allocation problem. Optimal closed-form expressions are derived for the inner problem and the Hungarian method is invoked to solve the outer problem. Numerical results demonstrate that our proposed approach achieves lowest AoI compared to the other baseline approaches.
\end{abstract}

\vspace{-0.2cm}
\section{Introduction}
Real-time Internet-of-Things (IoT) applications such as intelligent transportation systems (ITS), augmented reality, industry 4.0, possess stringent requirements of frequent and timely information updates to make critical decisions. 
 For example, in cooperative autonomous driving, which is an ITS application, information  updates such as speed and vehicle position, along with other sensory data must be timely disseminated to other vehicles so that the decisions such as changing lanes or merging maneuvers must be taken. Out-dated or stale information updates are highly undesirable for these applications as they may call forth unreliable or erroneous decisions~\cite{abd2019role}. 
 \par Age of Information (AoI) is a performance metric that offers a rigorous way to quantify the information freshness as compared to other performance metrics such as the delay or latency \cite{kaul2012real}. AoI can be defined as the elapsed time since the most recent delivered status message was generated \cite{kaul2012real}. Earlier attempts on AoI analysis mainly focused on orthogonal multiple access (OMA) schemes, where a single user equipment (UE) may transmit its information update on a particular resource block (time and/or frequency)~\cite{maatouk2019minimizing}\cite{9665756}. However, the massive connectivity and the high spectral efficiency requirements of IoT applications brought forward non-orthogonal multiple access (NOMA) as a promising multiple access scheme for future wireless networks
 against its contemporary OMA techniques~\cite{liu2020uplink}. NOMA empowers multiple UEs to transmit their information on the same resource block by adopting superposition coding (SC) at the transmitter side and by performing successive interference cancellation (SIC) at the receiver side \cite{kara2018error}. 
 \par Recent studies towards unleashing the full potentials of NOMA towards AoI optimization in real-time IoT applications unveiled that the impact of the wireless channels can severely impact the prompt delivery of information update messages  \cite{wang2021optimizing}. Generally, the highly random and uncontrollable behaviour of wireless communication environments hinders the in-time delivery of information updates. Typically, the blockages and channel impairments make it nearly impossible to attain a strong communication link between a source and a destination. Owing to reconfigurable intelligent surface (RIS), the propagation induced impairments of the wireless environments can be diminished and a strong channel can be potentially constructed between the source and the destination~\cite{9122596}.
Specifically, RIS consists of a planar array of passive elements, where each element is able to independently tune the phase-shift of the incident waves. Thus,  through a proper adjustment of the phase-shifts of all the RIS elements, signals can be strengthened at the points of interest \cite{Marco_JSAC}.
\par Due to its effectiveness, the integration of RIS in NOMA-based systems has recently drawn a significant attention from the research community. Precisely, the focus had been to explore the benefits of RIS in NOMA-based systems to maximize the sum data rate \cite{9500202}, to improve the network coverage range \cite{9462949}, to enhance the secrecy performance \cite{9385957}, and to boost the energy efficiency \cite{fu2019intelligent}. However, the proposed approaches may not be necessarily optimal neither for the freshness of information nor for uplink transmissions. Nevertheless, the works \cite{samir2020optimizing,muhammad2021age} have exploited the use of RIS for the AoI optimization. The authors in \cite{samir2020optimizing} studied how aerial-RIS can help to minimize  AoI for IoT applications. However, the authors assumed that only one IoT device (IoT) can be scheduled in each time-slot. Moreover, this work neglected the impact of the direct channels from the BS to the IoTDs. On the other hand, the authors in \cite{muhammad2021age} enhanced the AoI analysis by optimizing the RIS configuration for a finite number of IoTDs. where the impact of the direct channels from the BS to the IoTDs was also considered. Although, both of the discussed works signified the benefits of utilizing RIS to minimize the AoI, the scope is limited to OMA schemes. Recently, NOMA has been integrated with RIS for the first time in \cite{muhammad2022optimizing} to optimize the AoI, where a single RIS is considered to enhance the quality of the wireless channels. To the best of our knowledge, none of the previous works reported in the literature have considered the integration of multiple RISs with NOMA while considering the freshness of information, which thus motivates this work.
\par Against the above background, this paper aims to investigate the envisioned benefits of integrating multiple RISs in uplink NOMA systems, where multiple IoTDs are updating the base station (BS) by transmitting their time-sensitive information. Multiple RISs are deployed to enhance the quality of the wireless channels. For this setup, and with the objective of minimizing the average sum AoI, the phase-shift configurations of the RISs, the clustering of the IoTDs, and their associated transmit power are jointly optimized. The formulated problem is a mixed integer non-convex problem, which is difficult to solve. In order to address this issue, we first determine the assignment of RIS for the weak IoTDs followed by obtaining the phase-shift matrix for each RIS by maximizing the minimum channel gains of the weak IoTD. Finally, the problem is reformulated as a bi-level optimization problem, which is comprised of an inner power allocation problem and an outer user clustering problem. The inner problem is a feasibility condition problem for which the feasibility range for the transmit power of the IoTDs are derived. On the other hand, the outer problem is a classical linear assignment problem, which is solved through the Hungarian method.
\section{System Model}
\label{sysmodel}
\subsection{Network Model}
\begin{figure}[t]
\centering
\includegraphics[width=0.85\linewidth]{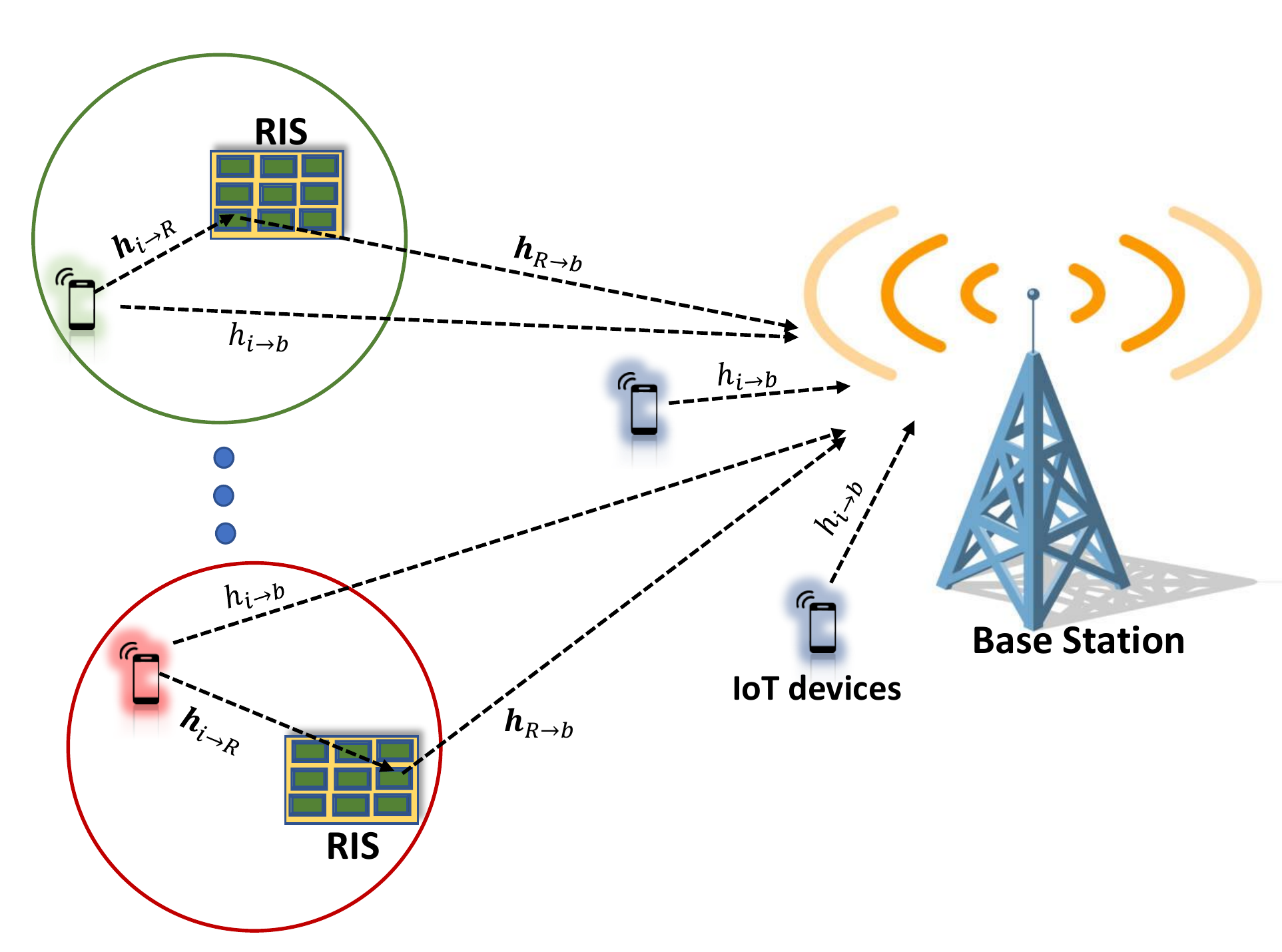}
\caption{An illustration of our system model.}
\label{systemmodel}
\end{figure}

As depicted in Fig. \ref{systemmodel}, we consider an RIS-empowered uplink IoT network that consists of one BS and $2I$ IoTDs ($I \in \mathbb{N}$). Let $[0,T]$ be the observation interval of the system, where $T$ represents the total observation duration. The observation interval is divided into time slots with a slot index $t = \{1, 2, \dots , T\}$. We assume that each IoTD and the BS are equipped with one transmit and one receive antenna~\cite{samir2020optimizing}. Let the set of IoTDs be denoted by $\mathcal{I} = \{1, 2, \dots ,  2I\}$. These IoTDs are divided into two disjoint sets with equal size $I$ based on their channel gains \cite{9140040}. The first set is denoted by $\mathcal{S}$ and contains the $I$ IoTDs that have the highest channel gains, which are referred to as strong IoTDs. The second set is denoted by $\mathcal{W}$ and contains, on the other hand, the $I$ IoTDs that have the lowest channel gains, which are referred to as weak IoTDs. The $2I$ IoTDs are grouped into $I$ disjoint clusters, where each cluster is formed by pairing exactly one IoTD from each set of IoTDs. Here, the uplink NOMA is adopted by each cluster to communicate with the BS. Furthermore, to abolish the inter-cluster interference at the BS, it is assumed that different clusters communicate with the BS simultaneously over orthogonal frequency resources \cite{9140040}. 
\par Owing to the impurities and obstacles of the wireless propagation environment, it is onerous to obtain a strong direct line-of-sight communication link between each weak IoTD and the BS. In this regard, a set of RISs denoted by $\mathcal{N} = \{1, 2, \dots ,  N\}$ is assumed to be deployed to enhance the uplink transmission from the weak IoTDs to the BS, where each RIS is equipped with $L$ reflecting elements. To maintain the needed quality of service (QoS) of the weak IoTDs, the BS continuously controls the phase shifts of the reflecting elements of each RIS. Let $ \boldsymbol \Phi_n(t)=  {\rm diag} \left( \exp \left[j \boldsymbol{\theta}^n(t)\right] \right) \in \mathbb{C}^{L \times L}$ be the diagonal phase shift matrix of the $n^\text{th}$ RIS at time-slot $t$, where $\boldsymbol{\theta}^n(t) = \left[\theta_1^n(t), \dots, \theta_L^n(t)\right]$ and,  $ \forall l \in L$, $\theta_l ^n(t) \in [0,2\pi)$ is the phase shift of the $l$th reflecting element of the $n^\text{th}$ RIS.  We assume that each weak IoTD $w \in \mathcal{W}$ is served by a set of RISs denoted by $ \Omega_w$, i.e.,   $1 \leq  |\Omega_w| \leq N$. Let $\boldsymbol \Omega = \{\Omega_1, \Omega_2, \dots, \Omega_I\}$ be the set of all RISs assignments.  
\par Literally, it seems a good solution to let all the IoTDs $w \in \mathcal{W}$  be served by the $N$ RISs. However, such a solution amplifies the cost of prodigious channel estimation complexity. In fact, if all RISs are allowed to serve all the IoTDs, the total number of required channel estimations is equal to the product of the number of IoTDs and the sum of the elements of all RISs. Therefore, the computational time of the channel estimation process becomes extremely high since the channel estimation process is a complex and time-consuming problem. On the other hand, if each  IoTD $w \in \mathcal{W}$ is restricted to be served by only one RIS to reduce the channel estimation complexity, the required QoS constraints may not be achieved. Based on this, a more flexible solution is required in order to avoid the two aforementioned extreme cases. Hence, the question that arises here is the following: \textit{what is the best assignment strategy that maximizes the benefits brought by the RISs to the weak IoTDs?}. 
\par To answer the above question. we assume in this paper that each weak IoTD is associated to $k$ RISs, with $1 \leq k \leq N$. Now, as per \cite{elhattab2021reconfigurable,9167258}, the users get the highest channel gains when they are located in close proximity to the RIS. Thus, each weak IoTD is assigned to its $k$ nearest RISs. 
\subsection{Channel Model}
In order to illustrate the signal model at the BS, we consider a cluster of IoTDs $(s,w) \in  \mathcal{S}\times\mathcal{W}$, where $s$ and $w$ represent the strong and weak IoTDs in the considered NOMA configuration, respectively. The channel gains between the RIS $n \in \mathcal{N}$ and the BS, between the weak user $w$ and the RIS $n \in \mathcal{N}$, and between IoTD $x \in \{s,w\}$ and the BS, are denoted by $\boldsymbol h_{R_n \rightarrow b}(t) \in \mathbb{C}^{L \times 1}$, ${\boldsymbol h_{w \rightarrow  R_{(w,n)}}}(t) \in \mathbb{C}^{L \times 1}$, and $h_{x \rightarrow b}(t)  \in \mathbb{C}$, respectively. We consider that all the channel gains consist of both the small-scale and the large-scale fading and we adopt the same models considered in  \cite{muhammad2022optimizing,elhattab2021reconfigurable}.
\subsection{AoI Modeling}
In the context of this paper, the AoI illustrates how fresh the information of an IoTD is from the BS's perspective and characterizes the inter-delivery time of arrivals and the latency \cite{kaul2012real}. The AoI of the $i$th IoTD at the BS in $t$th  time-slot is denoted as $\mathcal{Y}_{i}(t)$, for all $i \in \mathcal{I}$ and $t \in T$. A successful transmission at the BS in a given time slot $t$ occurs when the signal-to-interference-plus-noise-ratio (SINR) of the channels between the IoTDs $s$ and $w$ of the cluster $(s,w)$ $\in \mathcal{S}\times\mathcal{W}$
and the BS is above a given threshold denoted as $\zeta_{\rm{th}}$, which will drop down the AoI of each IoTD to one \cite{bhat2020throughput}. On the other hand, if the transmission remains unsuccessful, the AoI value will linearly increase by one \cite{bhat2020throughput}. This work considers a \textit{generate at will} model \cite{wang2021optimizing}, where a packet is always generated at the beginning of the time-slot and the transmission occurs in the same time-slot. Based on this, the evolution of AoI of $i$th IoTD is given by
\begin{equation}
 \mathcal{Y}_{i}({t}) = 
 \begin{cases}
  {1}, & \text{if  }    U(I-i)  \zeta_{s}(t) \text{+} \\
  & \,  U(i- I)  \zeta_{w}(t)  
  \geq \zeta_{\mathrm{th}}, 
    \\
  \mathcal{Y}_{i}({t-1})+1, &\text{otherwise,}
 \end{cases}
 \label{mainage}
\end{equation}
where $U(x) = 1$ if $x \geq 0$, and 0 otherwise.
 \section{Problem Formulation and Solution Approach}
 \subsection{Problem Formulation}
 Our aim is to minimize the average sum AoI over the entire observation interval $\left[0,T\right]$, which is given by 
 \begin{equation}
    \mathcal{Y}(T) = \dfrac{1}{T} \sum_{t<T} \Big[\dfrac{1}{2I} \sum_{s \in \mathcal{S}} \sum_{w \in \mathcal{W}} \Big\{  \mathcal{Y}_{s}(t) + \mathcal{Y}_{w}(t) \Big\} b_{s,w}(t) \Big],
\end{equation}
 In the context of our system model, since the generate-at-will model is adopted, a packet is assumed to be generated when it is requested by the BS. Based on this, the preset goal translates into minimizing the sum AoI at each time slot $t \in \{1, 2, \dots , T\}$ with respect to the strong and weak IoTDs. To do so, at each time slot$t \in \{1, 2, \dots , T\}$, we optimize the sum AoI with respect to the phase-shift matrices of the RISs $\boldsymbol \Phi= 
 [\boldsymbol\Phi_1,\boldsymbol\Phi_2,....,\boldsymbol\Phi_N]$, the transmit power of the strong IoTDs $\boldsymbol{p_s} = \{p_s | s \in \mathcal{S}\}$ and the weak IoTDs $\boldsymbol{p_w} = \{p_w | w \in \mathcal{W}\}$, and the clustering  policy $\boldsymbol B = \{ b_{s,w} \in \{0,1\} | (s,w)  \in \mathcal{S}\times\mathcal{W}\}$, where for all $(s,w)  \in \mathcal{S}\times\mathcal{W}$, $b_{s,w}(t)=1$ shows that IoTD $s \in \mathcal{S}$ is paired with IoTD $w \in \mathcal{W}$ at time-slot $t$, and $b_{s,w}(t)=0$, otherwise.
 \allowdisplaybreaks
\begin{subequations} 
\begin{align}
\text{$\mathcal {OP} $: } & \min_{ \substack{\boldsymbol B, \boldsymbol \Phi,\\ \boldsymbol{p_w}, \boldsymbol{p_s}}} \Big[\dfrac{1}{2I} \sum_{s \in \mathcal{S}} \sum_{w \in \mathcal{W}} \Big\{  \mathcal{Y}_{s}(t) + \mathcal{Y}_{w}(t) \Big\} b_{s,w}(t) \Big] \\
\text{s.t.  \,\,\, 
} & \label{constheta} \text{$\theta_l^n(t) \in [0,2\pi), \,\,\, \quad \,\, \,\,\,\,\forall \,\, l \in \mathcal{L}, n \in \mathcal{N}$,  } 
\\ 
&\label{consb1}  b_{s,w}(t)  \in \{0,1\}, \,\,\,\,\, \quad \forall w \in \mathcal{W}, s \in \mathcal{S}, \\
&\label{consbnf1} \sum\limits_{s=1}^ {I} b_{s,w}(t) \leq 1, \,\,\,\,\,\quad \forall w \in \mathcal{W}, \\
&\label{consbnf2} \sum_{w=1}^ {I} b_{s,w}(t) \leq 1, \,\,\,\,\, \quad \forall s \in \mathcal{S}, \\ 
&\label{consbnf3} p_w(t), p_s(t) \leq P_{\max}, \,\forall s \in \mathcal{S}, w \in \mathcal{W},
\end{align}
\end{subequations}
where $P_{\max}$ is the power budget at each IoTD. Constraints (\ref{consbnf1}) and (\ref{consbnf2}) guarantee that each strong IoTD $s \in \mathcal{S}$ is paired with at most one weak IoTD $w \in \mathcal{W}$ and  vice versa. Finally, constraint \eqref{consbnf3} ensures that each IoTD's transmit power does not exceed its power budget. It can be observed that $\mathcal{OP}$ is a  mixed-integer nonlinear programming problem due to the coexistence of the continuous variables ($\boldsymbol{\theta}^n, \boldsymbol{p}_w, \boldsymbol{p}_s$), and the binary clustering matrix $\boldsymbol{B}$, and hence,  it is difficult to solve it directly.
\vspace{-0.2cm}
\subsection{Solution Approach}
It can be observed from  $\mathcal{OP}$  that it is difficult to jointly obtain the optimal RIS phase-shift matrices, the optimal power control, and the optimal user clustering policy. Therefore, we resort to first obtaining the RIS phase-shift configurations that improves the channel gains of the weak IoTDs. Afterwards, the obtained RISs phase-shift matrices are injected into the original problem $\mathcal{OP}$ and the resulting problem turned out to be a joint user clustering and power control problem which is solved by leveraging the bi-level optimization. We now discuss each solution approach in details.
\subsubsection{Configurations of the RISs} 
As explained earlier, each weak IoTD is assigned in the beginning to its $k$ nearest RISs. The RIS phase-shift matrix optimization problem will be solved for each RIS, and is given as
\begin{subequations} 
\begin{align}
& \mathcal {OP}1:   \max_{ 
\boldsymbol \Phi_n} \min_{ 
 w  \in \mathcal{W}}    {| \boldsymbol h_{w \rightarrow R_{(w,n)}}(t) \boldsymbol \Phi_n(t) \boldsymbol h^H_{R_n \rightarrow b}(t) + h_{w \rightarrow b}(t) |^2}     \\
& \text{s.t.}  \quad \theta_l^n(t) \in [0,2\pi), \,\,\, \quad \,\, \,\,\,\,\forall \,\, l \in \mathcal{L},t \in T, n \in \mathcal{N}. \label{thetaf}
\end{align}
\end{subequations}
\noindent
Following the approach in \cite{muhammad2022optimizing}, by introducing an auxiliary variable $\xi$, problem $\mathcal {OP}1$ can be rewritten as
\begin{subequations} 
\begin{align}
&\hspace{-0.4cm}\mathcal{OP}2: \max_{ 
 \xi, \boldsymbol \Phi_n(t)} \,\,\,  \xi   \\
& \hspace{-0.4cm}\text{s.t.}\,\,\,\,\,\,\,\,\,\,
 \hspace{-0.4cm}\label{constheta2} | \boldsymbol h_{w \rightarrow R_{(w,n)}}(t) \boldsymbol \Phi_n(t) \boldsymbol h^H_{R_n \rightarrow b}(t) + h_{w \rightarrow b}(t) |^2 \geq \xi, \, \\
 & \,\,\, \quad \,\, \,\,\,\,\hspace{3cm} \forall w \in \mathcal{W}, t \in T, n \in \mathcal{N}, \notag\\
&\hspace{0.4cm}\theta_l^n(t) \in [0,2\pi), \,\,\, \quad \,\, \,\,\,\,\hspace{0.4cm}\forall \,\, l \in \mathcal{L},t \in T, n \in \mathcal{N}.
\end{align}
\end{subequations}
The above problem is transformed into a rank-one constrained optimization problem via change of variables and matrix lifting. Let $\boldsymbol u_n  \triangleq [ u_{n1},u_{n2},....,u_{nL}]^H$, where $u_{nl}=e^{j\theta_l^n}$ for all $l \in \mathcal{L}, n \in \mathcal{N}$. Thus, for all $l \in \mathcal{L}$ and $n \in \mathcal{N}$, the constraint $\theta_l^n(t) \in [0,2\pi)$ is equivalent to the unit-modulus constraints, i.e., $|u_{nl}|^2=$1. By applying the change of variables $\boldsymbol h_{w \rightarrow R_{(w,n)}}(t) \boldsymbol \Phi_n(t) \boldsymbol h^{H}_{R_n \rightarrow b}(t)  = \boldsymbol u^{H}_n  \boldsymbol{\mathcal{Z}}_n(t)$, where $\boldsymbol{\mathcal{Z}}_n(t) =\mathrm{diag}(\boldsymbol h^H_{R_n \rightarrow b}(t))\boldsymbol h_{w \rightarrow R_{(w,n)}}(t)$, we obtain $|\boldsymbol h_{w \rightarrow R_{(w,n)}}(t) \boldsymbol \Phi_n(t) \boldsymbol h^H_{R_n \rightarrow b}(t) + h_{w \rightarrow b}(t)|^2 = | \boldsymbol u^H_n \boldsymbol{\mathcal{Z}}_n(t) + h_{w \rightarrow b}(t)|^2 $ = $\bar {\boldsymbol u_n}^H  \boldsymbol{\Theta}_n \bar {\boldsymbol u_n} + |h_{w \rightarrow b}(t)|^2 = {\rm{tr}}(\boldsymbol{\Theta}_n \bar {\boldsymbol u_n} \bar {\boldsymbol u_n}^H) + |h_{w \rightarrow b}(t)|^2$, where  
\begin{align}
   \boldsymbol{\Theta}_n = \begin{bmatrix} \boldsymbol{\mathcal{Z}}_n(t)\boldsymbol{\mathcal{Z}}_n^H(t) & \boldsymbol{\mathcal{Z}}_n(t)h_{w \rightarrow b}(t) \\ h_{w \rightarrow b}(t)\boldsymbol{\mathcal{Z}}_n^H(t) & 0 \\ \end{bmatrix}, && \bar {\boldsymbol u_n}= \begin{bmatrix} \boldsymbol u_n  \\ 1 \end{bmatrix}.
\end{align}
\noindent Now, let  $\boldsymbol{\mathcal{U}}_n \triangleq {\boldsymbol u}_n \bar {\boldsymbol u}^H_n$, which needs to satisfy rank$(\boldsymbol{\mathcal{U}_n})$ = 1 and $\boldsymbol{\mathcal{U}_n} \geq$ 0. This rank one constraint is non-convex \cite{elhattab2021reconfigurable}. By dropping this constraint, problem $\mathcal{OP}2$ can be rewritten as
\begin{subequations} \label{eq:subeqf nscdglobalEQ2}
\begin{align}
\label{objop2}
\text{$\mathcal {OP} $3: } & \max_{ \boldsymbol{\mathcal{U}_n},
 \xi } \,\,\,  \xi   \\
&\text{s.t.}  \,\,\, tr(\boldsymbol{\Theta_n}\boldsymbol{\mathcal{U}_n}) + |h_{w \rightarrow b}(t)|^2 \geq  \xi, \\
& \text{$\boldsymbol{\mathcal{U}}_n \geq$ 0},
  \\
 &\label{vff} [\boldsymbol{\mathcal{U}}_n]_{L,L} = 1.
\end{align}
\end{subequations} 
 \par With the applied transformation, the resulting optimization problem appears to be a convex optimization problem that can be easily solved using any convex optimization solver such as CVX \cite{elhattab2021reconfigurable}. Note that, if the rank-one constraint is not satisfied, the Gaussian randomization (GR) method will be applied to construct a rank-one solution \cite{elhattab2021reconfigurable}. 
\begin{figure*}[t]
\vspace{0.04in}
\begin{align} 
  \zeta_{s}(t) &=
\frac{p_s(t) |h_{s \rightarrow b}(t) |^2} {p_w(t)|  h_{w \rightarrow b}(t) + \sum_{n \in    \mathcal{N}} h_{w \rightarrow R_{(w,n)} }(t) \boldsymbol \Phi_n (t) \boldsymbol h^H_{R_n  \rightarrow b}(t)  |^2+\sigma^2},
 \label{snrconstraintS} \tag{9}\\
 \zeta_{w}(t) & =
\frac{p_w(t)|  \sum_{n \in \Omega_w} h_{w \rightarrow R_{(w,n)} }(t) \boldsymbol \Phi_n (t) \boldsymbol h^H_{R_n  \rightarrow b}(t)  +\sum_{n' \in ( \boldsymbol\Omega \setminus \Omega_w)}   \boldsymbol h_{w \rightarrow R_{(w,n)} }(t) \boldsymbol \Phi_{n'} (t) \boldsymbol h^H_{R_n  \rightarrow b}(t) |^2}{\sigma^2}. \tag{10}
 \label{snrconstraintW}\\
 p_{w}^{\max} & = \min \left( \frac{ p_s|h^n_{s \rightarrow b}(t) |^2 - \zeta_{\rm th} \sigma^2 }{ \zeta_{\rm th}| \sum_{n \in \mathcal{N}} h_{w \rightarrow R_{(w,n)} }(t) \boldsymbol \Phi_n (t) \boldsymbol h^H_{R_n   \rightarrow b}(t)   |^2   }, P_{\max} \right), \label{eq:pwmax} \tag{11} \\
 p_{w}^{\min} & = \frac{\zeta_{\rm th} \sigma^2 }{|h_{w \rightarrow b}(t) + \sum_{n \in \Omega_w} h_{w \rightarrow R_{(w,n)} }(t) \boldsymbol \Phi_n (t) \boldsymbol h^H_{R_n  \rightarrow b}(t)  +\sum_{n' \in ( \boldsymbol\Omega \setminus \Omega_w)}   \boldsymbol h_{w \rightarrow R_{(w,n)} }(t) \boldsymbol \Phi_{n'} (t) \boldsymbol h^H_{R_n  \rightarrow b}(t) |^2}. \tag{12}
    \label{eq:pwmin}
  \end{align}
  \noindent\makebox[\linewidth]{\rule{\textwidth}{0.4pt}}
\end{figure*}
\subsubsection{SINR analysis and Bi-Level Optimization} 
\par In the considered RIS-empowered uplink NOMA system, the BS receives the signals from the IoTDs $s$ (through the direct link) and $w$ (through both the direct and the RIS(s) reflected links) in cluster $(s,w)$, respectively. Subsequently, SIC is sequentially applied to decode the signals of both IoTDs, where the decoding order is determined based on the channel gains of the IoTDs. Actually, the BS first decodes the signal of the strong IoTD $s$ whereas the signal of the weak IoTD $w \in \mathcal{W}$ is considered as an interference. As a result, the received SINR of IoTD $s$ at the BS in time-slot $t \in T$ can be expressed as \eqref{snrconstraintS} on top of next page, where $\sigma^2$ is the variance of zero-mean additive Gaussian noise and $p_s(t)$ and $p_w(t)$ are the transmit power of the IoTD $s$ and $w$ at time-slot $t$, respectively. Once successfully decoded, the signal of the strong IoTD $s$ is removed from the received signal at the BS and, then, the signal of IoTD $w$ will be decoded without any interference. The SINR of the IoTD $w$ at the BS in time-slot $t \in T$ is given by \eqref{snrconstraintW} on top of next page. It is worth mentioning that the BS receives the signal of the weak IoTD through the direct link, the assigned RISs to this weak IoT (first summation in the numerator of \eqref{snrconstraintW}), and the non-assigned RISs (second summation in the numerator of \eqref{snrconstraintW}). 
\par Assuming that the RISs configurations are obtained and served as an input to problem $\mathcal{OP}$, we can observe from $\mathcal{OP}$ that the AoI function is independent from the pairing variable $\boldsymbol B$. In fact, it is only a function of the power control policy as shown in \eqref{mainage}. Let the set of optimal user clustering and power control scheme, i.e., the solutions of the problem  $\mathcal{OP}$, be denoted by $\left\{\left(b^*_{s,w}(t),  p^*_s(t), p^*_w(t)\right),\,\, s,w \in \mathcal{S} \times \mathcal{W} \right\}$. For all $s,w \in \mathcal{S} \times \mathcal{W}$, if $b^*_{s,w}(t)=0$, then  $\left(p^*_s(t)),p^*_w(t)\right) = (0,0)$. However, if $b^*_{s,w}(t)=1$, then $\left(p^*_s(t)), p^*_w(t)\right)$ should be the optimal solutions for the power control policy of users $s$ and $w$.  To simplify the SIC process at the BS, the strong IoTD is considered to transmit at the maximum power $P_{\max}$. However, we still need to optimize the transmit power for weak IoTD. 
Therefore, for all $(s,w) \in \mathcal{S} \times \mathcal{W}$, if we assume that the IoTDs are paired together and the optimal power allocation is $p^*_w(t)$, then the problem $\mathcal{OP}$ ended up as a linear assignment problem that determines the optimal pairing policy $(b^*_{s,w})_{(s,w) \in \mathcal{S} \times \mathcal{W}}$. Here, for all $(s,w) \in \mathcal{S} \times \mathcal{W}$, the aim is to determine the optimal power allocation that satisfies the SINR threshold for each pair of IoTDs such that the AoI is minimized. Therefore, the feasible set of the power control of problem $\mathcal{OP}$ can be reduced to the set of the power allocation that minimizes the sum of AoI of each pair of IoTDs. Precisely, problem $\mathcal{OP}$ can be re-written as:
\setcounter{equation}{12}
\begin{subequations} 
\begin{align}
\text{$\mathcal {OP}_{outer} $: } & \min_{ \boldsymbol B}\Big[\dfrac{1}{2I} \sum_{s \in \mathcal{S}} \sum_{w \in \mathcal{W}} \Big\{  \mathcal{Y}_{s}(t) + \mathcal{Y}_{w}(t) \Big\} b_{s,w}(t) \Big] \\
& \text{s.t.  \,\,\, (\ref{consb1}),
(\ref{consbnf1}), (\ref{consbnf2}), 
}  
\end{align}
\end{subequations}
where $p^*_{w}(t)$ can be obtained by solving the following feasibility check optimization problem
\begin{subequations} 
\begin{align}
\hspace{-0.2cm}\mathcal{O}&\mathcal{P}_{inner}: \mathrm{Find}~~ \boldsymbol{p}_w \\
&\text{s.t.} \,\,\, p_w(t), p_s(t) \leq P_{\max}, \qquad \forall w \in \mathcal{W}, s \in \mathcal{S}, t \in T, \\
&\zeta_{s}(t) \geq \zeta_{\rm th}, \hspace{2.2cm}\,\,\,\,\,\,  \forall s \in \mathcal{I}, t \in T,\\
&\zeta_{w}(t) \geq \zeta_{\rm th}, \hspace{2.2cm}\,\,\,\,\,  \forall w \in \mathcal{I}, t \in T,
 \end{align}
\end{subequations}
\noindent for each pair of IoTDs $(s,w)  \in \mathcal{S} \times \mathcal{W}$. Accordingly, $\mathcal {OP}_{inner} $ is a power allocation problem for a given cluster of IoTDs and it defines the set of feasible solutions for problem $\mathcal {OP}_{outer}$, which is a linear assignment problem. Nonetheless, we need to solve $\mathcal {OP}_{inner}$ for all possible combinations of IoTDs $\left(s,w\right) \in \mathcal{S} \times \mathcal{W}$. Therefore we now present a computationally efficient approach to solve $\mathcal {OP}_{inner}$. 
\subsubsection{Power Control} Considering a pair of NOMA IoTDs $(s,w) \in \mathcal{S} \times \mathcal{W}$, our goal is to obtain a possible value of the power allocation for the weak user $w$, i.e., $p_{w}$, that satisfies the SINR constraints of IoTDs $s$ and $w$ when paired together. 
\begin{equation}
\zeta_{s}(t) \geq \zeta_{\rm th}~\mathrm{and}~\zeta_{w}(t) \geq \zeta_{\rm th}.
\label{eq:SINR_constraints}
\end{equation}
From the SINRs expressions in \eqref{snrconstraintS} and \eqref{snrconstraintW}, it can be concluded that the SINRs constraints in \eqref{eq:SINR_constraints} are satisfied if and only if 
\begin{equation}
    p_{w}^{\min} \leq  p_{w}^{\max},
\end{equation}
where  $p_{w}^{\max}$ and $p_{w}^{\min}$ are expressed  by \eqref{eq:pwmax} and \eqref{eq:pwmin} respectively.
\noindent
Consequently, any value of the power $p_{w}$ within the range $[p_{w}^{\min},p_{w}^{\max}]$ is a feasible value for $\mathcal{OP}_{inner}$.

\subsubsection{User Clustering} 

Once the optimal power allocation has been obtained for each possible NOMA cluster along with its corresponding sum AoI,  \textit{Hungarian method} can be utilized to determine the optimal clustering configuration \cite{9140040}. The core idea is that the \textit{Hungarian method} can optimally solve the 2-dimensional matching problem. Precisely, the input of the Hungarian method is an $I\times I$ cost matrix, which is nothing but the sum of AoI of the strong and weak IoTDs in each cluster, and its output is the pairing matrix $\boldsymbol B^*$ where $\boldsymbol B^*(s,w)$ = 1 shows that strong user $s \in \mathcal{S}$ is paired with weak user $w \in  \mathcal{W}$, and $\boldsymbol B^*(s,w)$ = 0 otherwise.  

\subsubsection{Complexity Analysis}
We discuss here the computational complexity of the proposed solution. After obtaining the closed form expression of the power, the computational complexity of obtaining the total transmit power is approximately $\mathcal{O}(1)$. On the other hand, the phase shift matrix problem is a semi-definite programming problem whose complexity along with applying Gaussian randomization is approximately $\mathcal{O}(\text{log}(1/\epsilon)(L^{4.5} + xT_{GR}))$ \cite{elhattab2021reconfigurable}.

\begin{figure*}[ht]
\centering
\subfigure[Impact of number of elements  per RIS.] {\centering\includegraphics[width=0.32\textwidth]{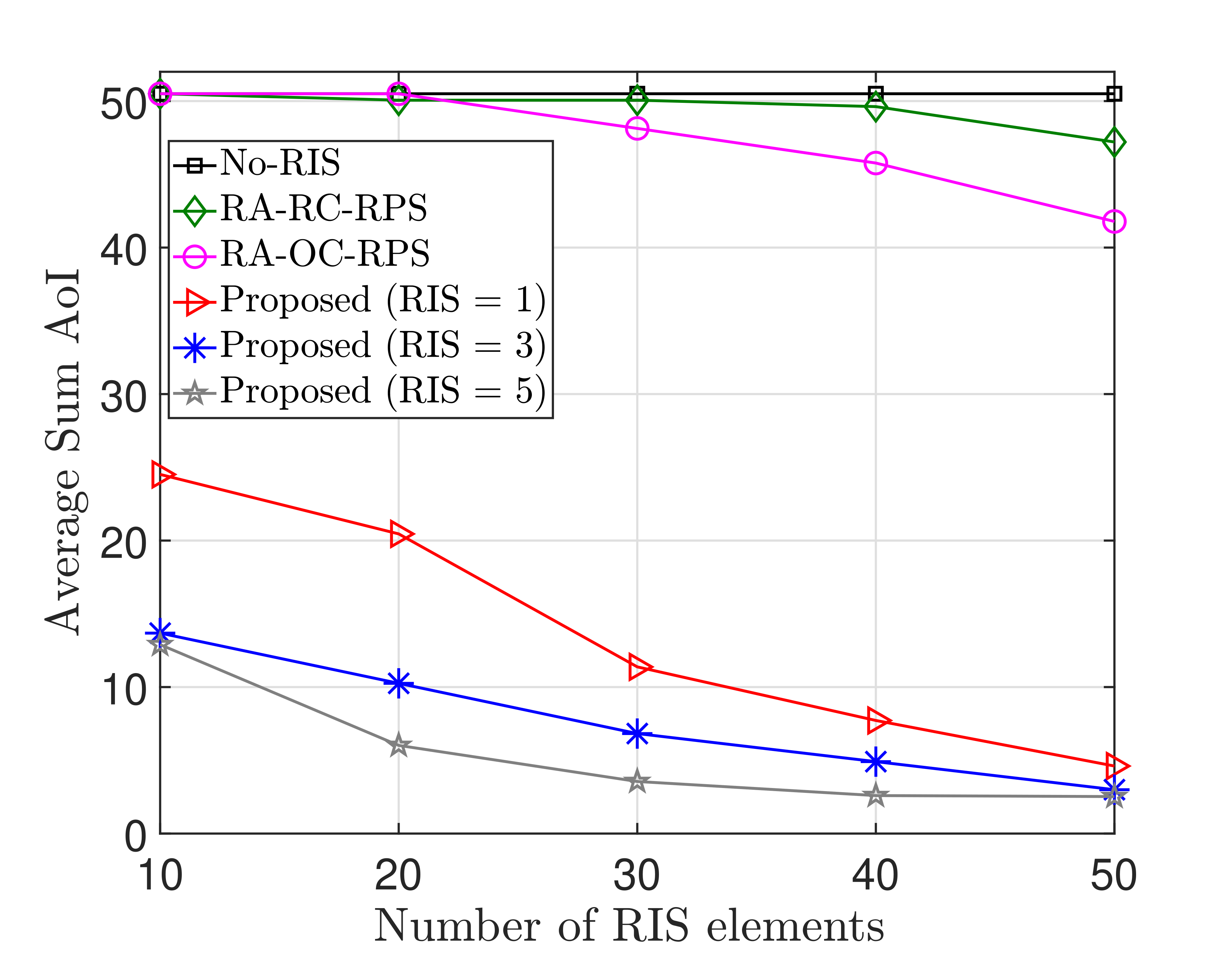}}
\subfigure[Impact of \# assigned RISs on AoI.]
{\centering\includegraphics[width=0.32\textwidth]{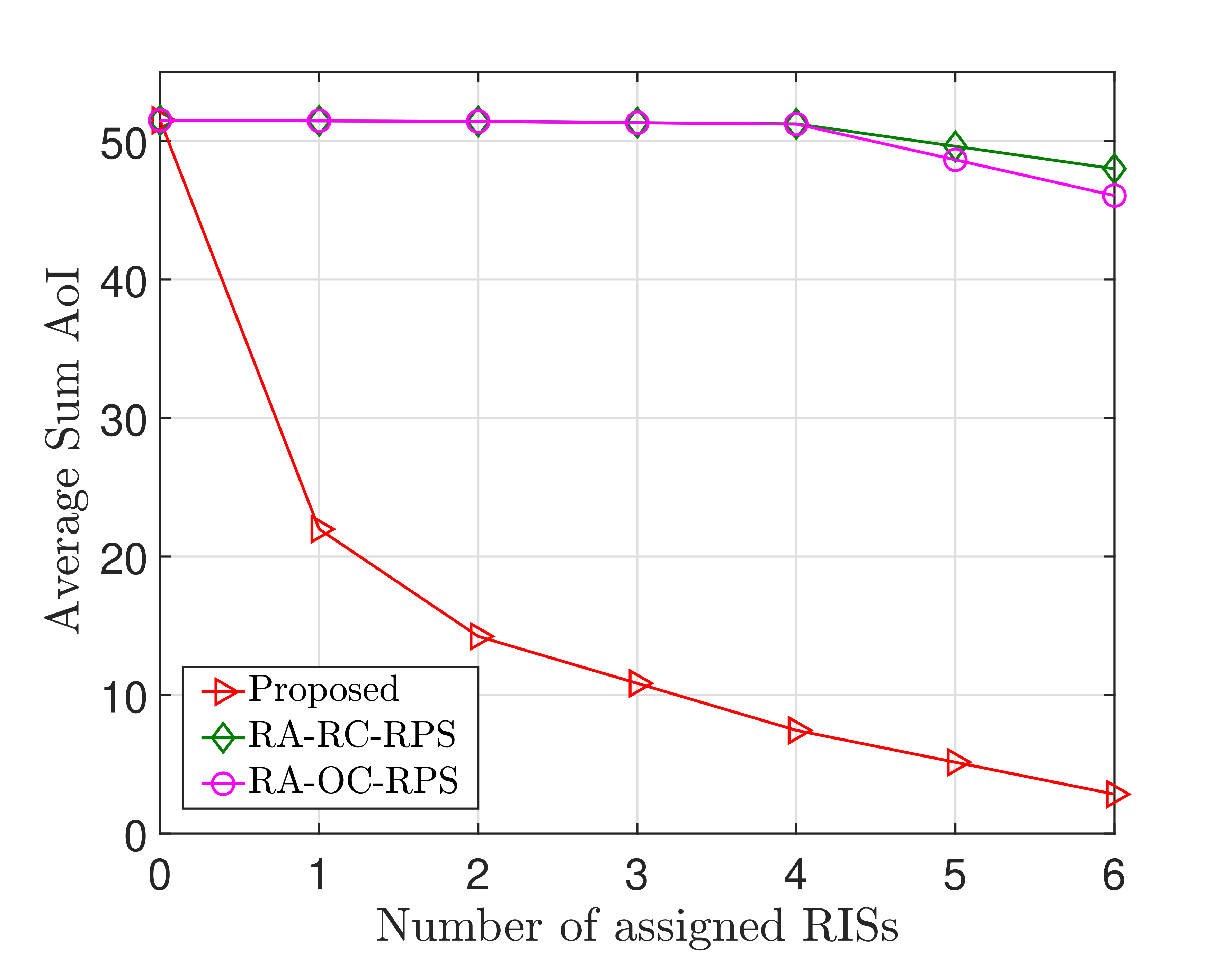}}
\subfigure[Impact of Power on AoI.] {\centering\includegraphics[width=0.32\textwidth]{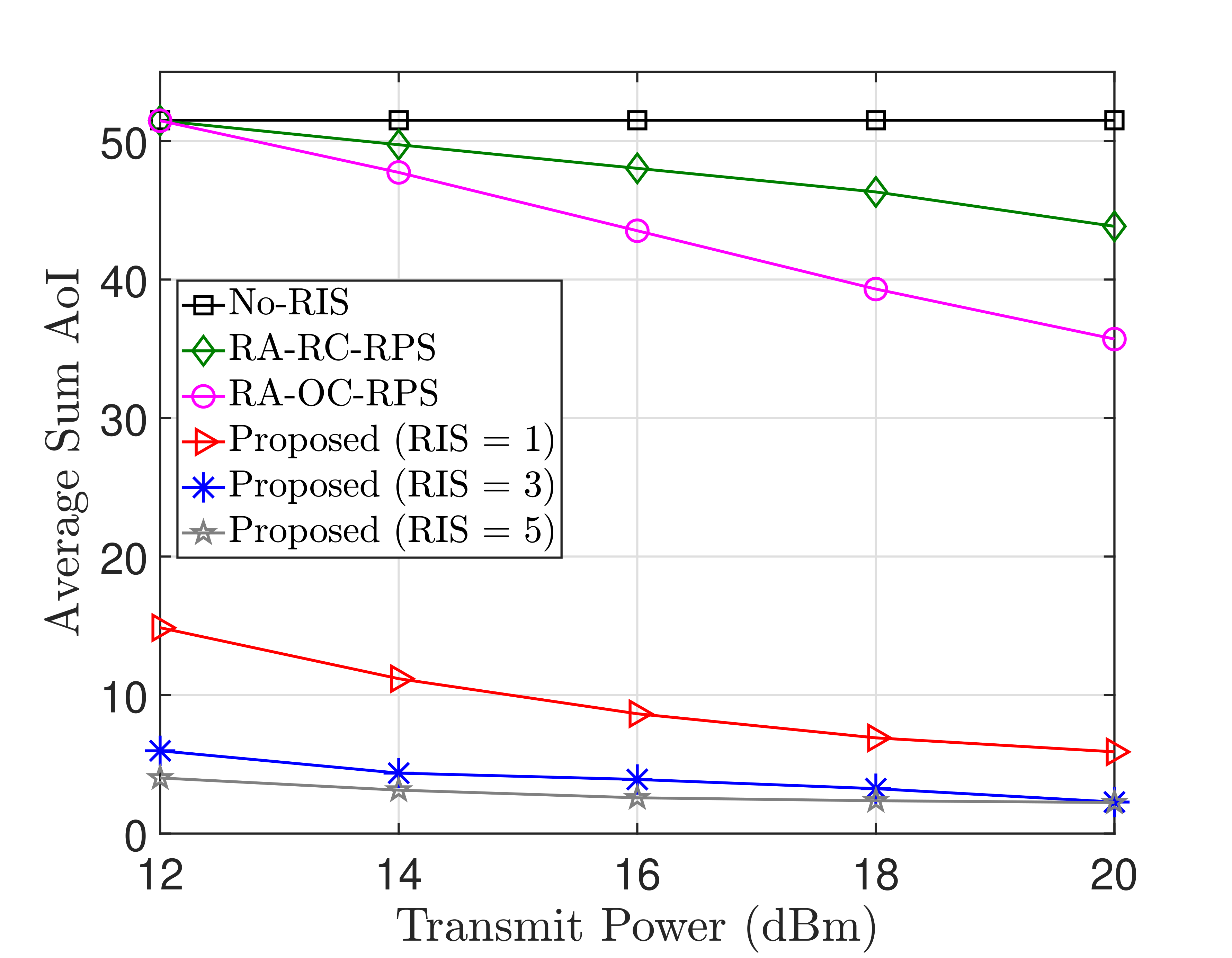}} 
\caption{Performance Evaluation}
\label{results}
\end{figure*}
 \section{Simulation and Numerical Analysis}
  \label{simulationresults}
\subsection{Simulation Settings}
We consider a 3-D area that consists of one BS, $N$ RISs and a set of IoTDs. The global coordinate system $(X, Y, Z)$ is Cartesian. The BS is located at $(0,0,H_{\rm b})$ and each RIS is located at $(x_n,y_n,H_{\rm r})$, where $H_{\rm b}$ and $H_{\rm r}$ are the heights of the transmit antenna of the BS and that of RIS, respectively. Multiple IoTDs are randomly distributed at the ground level, where for all $i \in \mathcal{I}$, the locations of IoTDs are $(x_i,y_i,0)$. There are two sets of IoTDs, i.e., strong IoTDs $\mathcal{S}$ and weak IoTDs $\mathcal{W}$. The strong IoTDs, weak IoTDs,  and the RISs are randomly located in a region $D_\mathcal{S}$, $D_\mathcal{W}$, and $D_\mathcal{R}$ with radius $d^s$, $d^w$, $d^r$ (i.e., $d^s < d^w < d^r$) and a central angle  $\Delta$ around the BS, similar to the distance model given in \cite{khazali2021energy}. 
The total number of time-slots $T=100$, the number of RISs $N =5$, and the number of IoTDs $I=20$, while the communication parameters are taken as $d^s_i=5$m, $d^w_{i}=145$m, $d^{r} =150$m, $\sigma^2= -110$ dBm,  $\eta_{R \rightarrow b} = \eta_{w \rightarrow R}= 2.2$, $\eta_{s \rightarrow b}$= $\eta_{w \rightarrow b}$ = 3.5, $K1$ = 2, $H_{\rm b}$=  $H_{\rm r}=10$m, $P_{max}=12$dBm, $\zeta_{\mathrm{th}}=45$dB.  
To assess the performance of our proposed approach, we have compared it with  other schemes. In the first scheme (denote as RA-OC-RPS), instead of assigning RISs that are in close proximity to the IoTDs, a random assignment has been utilized along with obtaining a random phase shift matrix. However, the optimal clustering is obtained. The second scheme (denote as  RA-RC-RPS) entails random RIS assignment along with a random clustering and random phase shift matrix. Moreover, we also looked at the impact of different number of assigned RISs on the average sum AoI, i.e., (RIS=1) indicates that each IoTD $w \in \mathcal{W}$ has been assigned to a single RIS, (RIS=3) indicates that each IoTD has been assigned to three RISs and so on. 
\subsection{Results and Discussions}
Fig.~\ref{results}(a) depicts the impact of the size of the RISs on the average sum AoI. The total number of RISs are set to five in this experiment whereas each IoTD is served by one, three and five RISs. It can be seen that the RIS has a significant impact on the AoI and increasing the number of the RIS elements results in decreasing the AoI. This is evident since the RIS helps to improve the channel quality of the weak IoTDs which leads to augmenting the likelihood of pairing with strong IoTDs and eventually ends up decreasing the AoI. In addition, we observe that the proposed algorithm achieves the lowest sum AoI compared to the other approaches. Particularly, the proposed scheme achieves around $94.43\%$, $92.8\%$, and $93.65\%$ decrease in the average AoI when three RISs (each having $50$ RIS elements) are assigned to each weak IoTD as compared to the No-RIS, RA-OC-RPS and RA-RC-RPS case 
respectively. We also note that if the number of RIS elements are small, IoTDs should be assigned to more than one RISs to achieve a lower average sum AoI. Moreover, it can be observed that  having three assigned RISs gets very similar AoI reduction as of assigning five RISs with large number of RIS elements, i.e., 50. Note that this insight could help network operators to decide on the number of RISs to be deployed in the network with their respective sizes in terms of RIS elements.   
\par Fig.~\ref{results}(b) illustrates the average sum AoI that is achieved by varying the number of RISs assigned to each weak IoTD. As can be observed, the average sum AoI attains its maximum value  when no RIS has been utilized. As the number of assigned RISs is increased, the average sum AoI starts declining. We can see a sharp decrease in the AoI curve (around 57.4\% decrease) when a single RIS is assigned as compared to no-RIS assignment case. However, the gap in AoI reduction starts reducing as more number of RISs are assigned (e.g., around 24.4\% decrease when the number of assigned RISs increases from 5 to 6). Moreover, both the other schemes with random assignment and random phase shift matrix attains the highest AoI. 
\par Fig.~\ref{results}(c) plots the average sum AoI against the maximum transmit power per IoTD. The total number of RISs are kept as five in this experiment with each RIS having $30$ elements. It is demonstrated that increasing the power directly increases the SINR, and thus, results in reducing the AoI. Nevertheless, our proposed approach outperforms others to a large extent. Moreover, the importance of the optimal clustering when relying on random RIS configuration can also be noted. The AoI achieved by the random clustering gets around $18.56\%$ higher than the one with the optimal clustering with power budget of each user $= 20$dBm. Finally, it always pays off to increase the number of RIS assignment. However, it can be compensated with the increase  of the IoTDs transmit power. 
 \section{Conclusion}
In this paper, the integration of RIS and NOMA in uplink IoT networks has been investigated to optimize the freshness of information. An optimization problem has been formulated to minimize the average sum AoI by optimizing the user-clustering, user transmit power and the RIS configuration. The RIS configuration problem is solved using the SDR approach. Then, the joint user-clustering and power control problem is decomposed into disjoint user-clustering and power-control. The Hungarian method is adopted to solve the user-clustering sub-problem while the feasible transmit power range is obtained for the power-control sub-problem. Numerical results demonstrated that the proposed method has superior performance in terms of achieving the lowest AoI.

\label{conclusion}
 \bibliographystyle{IEEEtran}
 \bibliography{IEEEabrv,main.bib}

\begin{thebibliography}{10}
\providecommand{\url}[1]{#1}
\csname url@samestyle\endcsname
\providecommand{\newblock}{\relax}
\providecommand{\bibinfo}[2]{#2}
\providecommand{\BIBentrySTDinterwordspacing}{\spaceskip=0pt\relax}
\providecommand{\BIBentryALTinterwordstretchfactor}{4}
\providecommand{\BIBentryALTinterwordspacing}{\spaceskip=\fontdimen2\font plus
\BIBentryALTinterwordstretchfactor\fontdimen3\font minus
  \fontdimen4\font\relax}
\providecommand{\BIBforeignlanguage}[2]{{%
\expandafter\ifx\csname l@#1\endcsname\relax
\typeout{** WARNING: IEEEtran.bst: No hyphenation pattern has been}%
\typeout{** loaded for the language `#1'. Using the pattern for}%
\typeout{** the default language instead.}%
\else
\language=\csname l@#1\endcsname
\fi
#2}}
\providecommand{\BIBdecl}{\relax}
\BIBdecl

\bibitem{abd2019role}
Abd-Elmagid \emph{et~al.}, ``{On the role of age of information in the Internet
  of Things},'' \emph{IEEE Commun. Magazine}, vol.~57, no.~12, pp. 72--77,
  2019.

\bibitem{kaul2012real}
S.~Kaul \emph{et~al.}, ``{Real-time status: How often should one update?}'' in
  \emph{Proceedings IEEE INFOCOM}, 2012, pp. 2731--2735.

\bibitem{maatouk2019minimizing}
A.~Maatouk \emph{et~al.}, ``{Minimizing the age of information: NOMA or OMA?}''
  in \emph{IEEE INFOCOM WKSHPS}, 2019, pp. 102--108.

\bibitem{9665756}
A.~Muhammad \emph{et~al.}, ``{Minimizing Age of Information in Multi-Access
  Edge Computing-assisted IoT Networks},'' \emph{IEEE Internet Things J.}, pp.
  1--1, 2021.

\bibitem{liu2020uplink}
X.~Liu \emph{et~al.}, ``{Uplink resource allocation for NOMA-based hybrid
  spectrum access in 6G-enabled cognitive Internet of Things},'' \emph{IEEE
  Internet Things J.}, 2020.

\bibitem{kara2018error}
F.~Kara \emph{et~al.}, ``{On the error performance of cooperative-NOMA with
  statistical CSIT},'' \emph{IEEE Communications Letters}, vol.~23, no.~1, pp.
  128--131, 2018.

\bibitem{wang2021optimizing}
Q.~Wang \emph{et~al.}, ``{Optimizing information freshness via multiuser
  scheduling with adaptive {NOMA/OMA}},'' \emph{IEEE Trans. Wireless Commun.},
  2021.

\bibitem{9122596}
S.~{Gong} \emph{et~al.}, ``{Toward Smart Wireless Communications via
  Intelligent Reflecting Surfaces: A Contemporary Survey},'' \emph{IEEE Commun.
  Surveys Tuts.}, vol.~22, no.~4, pp. 2283--2314, 2020.

\bibitem{Marco_JSAC}
M.~Di~Renzo \emph{et~al.}, ``{Smart Radio Environments Empowered by
  Reconfigurable Intelligent Surfaces: How It Works, State of Research, and The
  Road Ahead},'' \emph{IEEE J. Sel. Areas Commun.}, vol.~38, no.~11, pp.
  2450--2525, 2020.

\bibitem{9500202}
Z.~Yang \emph{et~al.}, ``{Machine Learning for User Partitioning and Phase
  Shifters Design in RIS-Aided NOMA Networks},'' \emph{IEEE Trans. Commun.},
  vol.~69, no.~11, pp. 7414--7428, 2021.

\bibitem{9462949}
C.~Wu \emph{et~al.}, ``{Coverage Characterization of STAR-RIS Networks: NOMA
  and OMA},'' \emph{IEEE Commun. Lett.}, vol.~25, no.~9, pp. 3036--3040, 2021.

\bibitem{9385957}
Z.~Zhang \emph{et~al.}, ``{Improving Physical Layer Security for Reconfigurable
  Intelligent Surface Aided NOMA 6G Networks},'' \emph{IEEE Trans. Veh.
  Technol.}, vol.~70, no.~5, pp. 4451--4463, 2021.

\bibitem{fu2019intelligent}
M.~Fu \emph{et~al.}, ``{Intelligent reflecting surface for downlink
  non-orthogonal multiple access networks},'' in \emph{IEEE Globecom}, 2019.

\bibitem{samir2020optimizing}
M.~Samir \emph{et~al.}, ``{Optimizing age of information through aerial
  reconfigurable intelligent surfaces: A deep reinforcement learning
  approach},'' \emph{IEEE Trans. Veh. Technol.}, vol.~70, no.~4, pp.
  3978--3983, 2021.

\bibitem{muhammad2021age}
\BIBentryALTinterwordspacing
A.~Muhammad \emph{et~al.}, ``{Age of Information Optimization in a
  {RIS}-Assisted Wireless Network},'' 2021. [Online]. Available:
  \url{https://arxiv.org/abs/2103.06405}
\BIBentrySTDinterwordspacing

\bibitem{muhammad2022optimizing}
\BIBentryALTinterwordspacing
A.~{Muhammad} \emph{et~al.}, ``{Optimizing Information Freshness in
  RIS-assisted NOMA-based IoT Networks},'' 2022. [Online]. Available:
  \url{https://arxiv.org/abs/2202.13572}
\BIBentrySTDinterwordspacing

\bibitem{9140040}
P.~Hũu \emph{et~al.}, ``{A Low-Complexity Framework for Joint User Pairing and
  Power Control for Cooperative NOMA in 5G and Beyond Cellular Networks},''
  \emph{IEEE Trans. Commun.}, vol.~68, no.~11, pp. 6737--6749, 2020.

\bibitem{elhattab2021reconfigurable}
M.~Elhattab \emph{et~al.}, ``{Reconfigurable intelligent surface enabled
  full-duplex/half-duplex cooperative non-orthogonal multiple access},''
  \emph{IEEE Trans. Wireless Commun.}, 2021.

\bibitem{9167258}
J.~Zuo \emph{et~al.}, ``{Resource Allocation in Intelligent Reflecting Surface
  Assisted NOMA Systems},'' \emph{IEEE Transactions on Communications},
  vol.~68, no.~11, pp. 7170--7183, 2020.

\bibitem{bhat2020throughput}
R.~V. Bhat \emph{et~al.}, ``{Throughput maximization with an average age of
  information constraint in fading channels},'' \emph{IEEE Trans. Wireless
  Commun.}, vol.~20, no.~1, pp. 481--494, 2020.

\bibitem{khazali2021energy}
A.~Khazali \emph{et~al.}, ``{Energy Efficient Uplink Transmission in
  Cooperative mmWave NOMA Networks with Wireless Power Transfer},'' \emph{IEEE
  Trans. Veh. Technol.}, 2021.

\end{thebibliography}
\end{document}